\documentclass[twocolumn,showpacs]{revtex4} %
\usepackage{epsfig,amsmath,amssymb,graphicx,verbatim}
\usepackage{natbib}
\usepackage[english]{babel} 

\begin{document}

\vspace{0mm}
\title{ On the dependence of the speed of light in vacuum on temperature } %
\author{Yu.M. Poluektov}
\email{yuripoluektov@kipt.kharkov.ua} %
\affiliation{National Science Center ``Kharkov Institute of Physics
and Technology'', Akhiezer Institute for Theoretical Physics, 61108
Kharkov, Ukraine}

\begin{abstract}
It is shown that the interaction of the electromagnetic field with
the vacuum of the electron-positron field gives rise to dependence
of the speed of light propagation on the radiation temperature. %
Estimates show that in the modern epoch, even at very high
temperatures, such for example which exist in the star interiors,
the temperature-dependent correction to the speed of light proves to
be extremely small. But in the cosmological model of the hot
Universe, in the first instances after the Big Bang the temperature
was so high that the speed of light exceeded its present value by
many orders of magnitude. The effect of dependence of the speed of
light on temperature must be important for understanding the early
evolution of the Universe.
\newline%
\end{abstract}
\maketitle

\section{Introduction}\vspace{-0mm} 

Maxwell's classical equations in vacuum are linear and contain a
fundamental constant of the velocity dimension which has the meaning
of the speed of propagation of electromagnetic waves. Nevertheless
within the framework of quantum electrodynamics the interaction of
the electromagnetic field with the vacuum of the electron-positron
field leads to interaction of photons with each other \cite{AB}.
Consequently, the equations of electromagnetic field become
nonlinear. Although this nonlinearity and the effects of scattering
of light on light are as a rule negligibly small, they can lead to
qualitatively new phenomena, in particular to dependence of the
speed of light of the equilibrium radiation on temperature. The
dependence of the speed of propagation of light in a material medium
on temperature is a natural effect, because the dielectric
permittivity of a medium depends on thermodynamic variables and, in
particular, on temperature. In the case under consideration in this
work, at issue is the dependence of the speed of light on
temperature in vacuum, by which the complex nature of the physical
vacuum is manifested. The temperature correction to the speed of
light was considered earlier in work \cite{Tarrach}. The dependence
of modification of vacuum, caused by different external conditions,
on the speed of propagation of light at finite temperature was
studied in works \cite{Scharnhorst,Barton,Latorre,Dittrich,Scharnhorst2}. %

In present work the influence of the interaction of photons on the
speed of their propagation is considered within the self-consistent
field theory, which allows to take into account the effect of the
mean field created by photons, whose magnitude considerably depends
on the number density of photons. In order to describe the
equilibrium electromagnetic field we will use the self-consistent
field model in the version which for the nonrelativistic Fermi and
Bose systems was developed in \cite{P1,P2,P3}. With regard to the
relativistic field models, this approach was applied in
\cite{P4,P5}. The influence of the nonlinear effects on the
propagation of phonons was investigated within this approach in
works \cite{P6,P7}. Therein, the Debye theory was generalized to
take into account the interaction of photons, and it was shown that
the speed of propagation of phonons increases with temperature.
Here, a similar approach is applied to describe a system of
interacting photons, whose interaction is accounted for by means of
the Euler-Heisenberg Hamiltonian \cite{AB}. It is shown that at low
frequencies, such that $\hbar\omega\ll m c^2$, where $\omega$ is the
photon frequency, $m$ is the electron mass, $c$ is the speed of
light, the correction to the speed of light is positive and
proportional to $T^4$. In this limit the considered effect is very
small. It is of interest to consider this effect in the opposite
high-frequency limit $\hbar\omega\gg m c^2$, which corresponds to
very high temperatures. Although, strictly speaking, in this limit
we go beyond the scope of applicability of description by means of
the Euler-Heisenberg Hamiltonian, nevertheless this model problem is
of notable interest and leads to reasonable results. In particular,
in this high-temperature limit all thermodynamic relations prove to
be satisfied, indicating a self-consistency of the used approach.

Temperatures, corresponding to this limit, are apparently cannot be
realized at modern conditions. But the condition $\hbar\omega\gg m
c^2$ had to be realized in the first instances of evolution of the
Universe after the Big Bang when temperatures were anomalously high
in comparison with temperatures of the modern epoch. In this early
stage of evolution the speed of light had to exceed the present one
by many orders of magnitude. In this work we calculate the
thermodynamic characteristics of the equilibrium radiation with
taking into account the dependence of the speed of light on
temperature. $$ $$ \vspace{50mm}

\section{Self-consistent description of the nonlinear electromagnetic field} 
Usually the Hamiltonian of a considered many-particle system can be
represented as a sum of the Hamiltonian of noninteracting particles
and the operator of their interactions. When using perturbation
theory, most often one chooses the Hamiltonian of noninteracting
particles as the main approximation and considers the operator of
their interaction as perturbation. But the effectiveness of
perturbation theory considerably depends on the successful choice of
the main approximation. The choice of the Hamiltonian of free
particles as the main approximation proves, as a rule, to be
unsuccessful, and to obtain physically correct results one needs to
sum an infinite number of terms\,\cite{AGD}. With decreasing
temperature the contribution of the kinetic energy into the total
energy of a system decreases and the interaction energy between
particles is coming to the foreground, as it no longer can be
considered as a small correction to the kinetic energy. Moreover,
the neglect of interaction in the main approximation does not allow
to study effectively the phase transitions. It is possible, however,
to reformulate perturbation theory to take the interaction into
account approximately already in the main approximation by the
self-consistent field \mbox{method \cite{P1,P2,P3,P4,P5}}.
Accounting for the phonon-phonon interaction in a solid in the
continuum Debye model \cite{P6,P7} by means of this method leads to
renormalization of the speed of sound and arising of its dependence
on temperature. In this work a similar approach is applied for
analysis of influence of the photon-photon interaction on light
propagation in vacuum.

The energy density of the electromagnetic field can be presented as
a sum of two terms
\begin{equation} \label{01}
\begin{array}{l}
\displaystyle{%
  w=w_0+w_I, %
}%
\end{array}
\end{equation}
where the first term which is quadratic in the electric and magnetic
field intensities
\begin{equation} \label{02}
\begin{array}{l}
\displaystyle{%
  w_0=\frac{{\bf E}^2+{\bf H}^2}{8\pi} %
}%
\end{array}
\end{equation}
determines the energy of the noninteracting electromagnetic field,
and the second term
\begin{equation} \label{03}
\begin{array}{l}
\displaystyle{%
  w_I=2D\left[3{\bf E}^2{\bf E}^2-{\bf H}^2{\bf H}^2-\left({\bf E}^2{\bf H}^2+{\bf H}^2{\bf E}^2\right)\right]+ %
}\vspace{2mm}\\ %
\displaystyle{\hspace{06mm}%
  +\,7D\left[({\bf E}{\bf H})^2+({\bf H}{\bf E})^2\right] %
}
\end{array}
\end{equation}
describes the interaction between photons due to creation of virtual
electron-positron pairs \cite{AB}. The constant in (\ref{03}) can be
calculated by the methods of quantum electrodynamics \cite{AB} and
in Gaussian units $\displaystyle{D\equiv\eta\frac{\hbar^3}{m^4c^5}}$, %
where the dimensionless coefficient
$\displaystyle{\eta\equiv\frac{\alpha^2}{45\,(4\pi)^2}}\approx 7.5\cdot\!10^{-9}$, %
$\alpha=e^2/\hbar c\approx 1/137$ is the fine-structure constant,
$m$  is the electron mass. The given coefficients contain the
constant $c$ having the dimension of speed, which we will call the
``bare'' speed of light. For estimation of coefficients the value of
this speed was taken equal to the observed speed of light, though,
as will be shown, it somewhat differs from the observed speed of
light at zero temperature. It is convenient to write the coefficient
in formula (\ref{03}) through the Compton wavelength of an electron
$\lambdabar=\hbar/mc$
in the form $\displaystyle{D=\eta\frac{\lambdabar^3}{mc^2}}$. %
It is interesting to estimate the value of the ratio of energies
$w_I/w_0$. This quantity is equal to the ratio of the field energy
contained in the volume $\lambdabar^3$ to the rest energy of an
electron. In addition, this ratio should be multiplied by the small
dimensionless coefficient $\eta$. For the magnetic field intensity
of the order of $H\sim 10^6$\,Gs we have $w_I/w_0\sim 10^{-20}$, so
that the contribution of interaction into the total field energy is
indeed extremely small.

Let us proceed to the description of the electromagnetic filed in
terms of the Fourier components of the fields, using the expansion
of the fields in plane waves
\begin{equation} \label{04}
\begin{array}{l}
\displaystyle{%
  {\bf E}({\bf r},t)=\sum_{k}{\bf E}_k(t)e^{i{\bf k}{\bf r}},\quad  %
  {\bf H}({\bf r},t)=\sum_{k}{\bf H}_k(t)e^{i{\bf k}{\bf r}}.
}
\end{array}
\end{equation}
Then the full Hamiltonian of the field in the volume $V$, in
accordance with (\ref{01}), is a sum of the free Hamiltonian and the
interaction Hamiltonian
\begin{equation} \label{05}
\begin{array}{l}
\displaystyle{%
  H=H_0+H_I, %
}
\end{array}
\end{equation}
where
\begin{equation} \label{06}
\begin{array}{l}
\displaystyle{%
  H_0=\frac{V}{8\pi}\sum_{k}\left({\bf E}_k^+{\bf E}_k+{\bf H}_k^+{\bf H}_k\right),  %
}
\end{array}
\end{equation}
\begin{equation} \label{07}
\begin{array}{l}
\displaystyle{\hspace{00mm}%
  H_I=2VD\times
}\vspace{2mm}\\ %
\displaystyle{\hspace{00mm}%
  \times\sum_{\{k_i\}}\Big\{3\!\left({\bf E}_{k_1}^+{\bf E}_{k_2}\right)\!\left({\bf E}_{k_3}^+{\bf E}_{k_4}\right) %
  -\!\left({\bf H}_{k_1}^+{\bf H}_{k_2}\right)\!\left({\bf H}_{k_3}^+{\bf H}_{k_4}\right)- %
}\vspace{0mm}\\ %
\displaystyle{\hspace{12mm}%
  -\left({\bf E}_{k_1}^+{\bf E}_{k_2}\right)\!\left({\bf H}_{k_3}^+{\bf H}_{k_4}\right) %
  -\!\left({\bf H}_{k_4}^+{\bf H}_{k_3}\right)\!\left({\bf E}_{k_2}^+{\bf E}_{k_1}\right)\!\Big\}\times %
}\vspace{2mm}\\ %
\displaystyle{\hspace{00mm}%
  \times\Delta\big({\bf k}_1-{\bf k}_2+{\bf k}_3-{\bf k}_4\big)\,+ %
}\vspace{3.5mm}\\ %
\displaystyle{\hspace{00mm}%
  +\,7VD\!\sum_{\{k_i\}}\!\Big\{\!\!\left({\bf E}_{k_1}^+{\bf H}_{k_2}\right)\!\left({\bf E}_{k_3}^+{\bf H}_{k_4}\right) %
  \!+\!\left({\bf H}_{k_4}^+{\bf E}_{k_3}\right)\!\left({\bf H}_{k_2}^+{\bf E}_{k_1}\right)\!\!\Big\}\times %
}\vspace{0mm}\\ %
\displaystyle{\hspace{00mm}%
  \times\Delta\big({\bf k}_1-{\bf k}_2+{\bf k}_3-{\bf k}_4\big).
}
\end{array}
\end{equation}
Here $\Delta\big({\bf k}\big)=1$ if ${\bf k}=0$ and %
$\Delta\big({\bf k}\big)=0$ if ${\bf k}\neq 0$. %
In (\ref{06}) and (\ref{07}) we can pass to the operators of
creation $a_{kj}^+$ and annihilation $a_{kj}$ of photons, using
representations of the operators of the Fourier components of the
fields:
\begin{equation} \label{08}
\begin{array}{l}
\displaystyle{%
  {\bf E}_k=-i\sqrt{\frac{2\pi\hbar\omega_k}{V}}\sum_{j}\!\big(a_{kj}^+-a_{-kj}\big)\,{\bf e}_j({\bf k}), %
}\vspace{2mm}\\ %
\displaystyle{\hspace{00mm}%
  {\bf H}_k=ic\sqrt{\frac{2\pi\hbar}{V\omega_k}}\sum_{j}\!\big(a_{kj}^++a_{-kj}\big)\big[{\bf k}\times{\bf e}_j({\bf k})\big], %
}
\end{array}
\end{equation}
where $\omega_k=ck$, and the polarization vectors ${\bf e}_j({\bf k})$ %
satisfy the conditions of orthonormality and completeness:
\begin{equation} \label{09}
\begin{array}{l}
\displaystyle{%
  {\bf e}_{j_1}^*({\bf k})\,{\bf e}_{j_2}({\bf k})=\delta_{j_1j_2},\,\,\, %
  \sum_j{{\bf e}_{j}^\alpha}^*({\bf k})\,{\bf e}_{j}^{\alpha'}({\bf k})=\delta_{\alpha\alpha'}-\!\frac{k_\alpha k_{\alpha'}}{k^2},%
}
\end{array}
\end{equation}
as well as the conditions
\begin{equation} \label{10}
\begin{array}{l}
\displaystyle{%
  {\bf k}\,{\bf e}_j({\bf k})=0, \qquad  {\bf e}_j^*(-{\bf k})={\bf e}_j({\bf k}).  %
}
\end{array}
\end{equation}
The free Hamiltonian of the field (\ref{06}) is reduced to a sum of
the Hamiltonians of harmonic oscillators
\begin{equation} \label{11}
\begin{array}{l}
\displaystyle{%
  H_0=\sum_{k,j}\hbar\omega_k\left(a_{kj}^+a_{kj}+\frac{1}{2}\right). %
}
\end{array}
\end{equation}
The electromagnetic field with account of the nonlinear effects is
characterized by the full Hamiltonian (\ref{05}). In order to
account for interaction in a many-particle system, usually one
chooses the Hamiltonian of noninteracting particles as the main
approximation and considers the interaction Hamiltonian as
perturbation (in our case, those are $H_0$ (\ref{11}) and $H_I$ (\ref{07})). %
Such choice, as remarked above, is not optimal, because the effects
caused by interaction are totally disregarded in the main
approximation. Although the interaction is small in the considered
case, it can lead, as we will see, to qualitatively new effects. It
is known from the self-consistent approach for description of
many-particles systems that accounting for the interaction effects
in the main approximation leads to a change in the dispersion law of
the initial particles and, thereby, we pass from the representation
of free particles to the language of collective excitations --
quasiparticles.

It is natural to consider that also in the case studied here the
interaction effects will lead to renormalization of the ``bare''
speed of light $c$ entering into the Hamiltonian. Taking into
account this consideration, let us decompose the full Hamiltonian
(\ref{05}) into the main part and the perturbation in a different
way, that is
\begin{equation} \label{12}
\begin{array}{l}
\displaystyle{%
  H=H_S+H_C, %
}
\end{array}
\end{equation}
where the self-consistent (or approximating) Hamiltonian is chosen
in the form similar to the free Hamiltonian (\ref{11}), but with the
speed of light $\tilde{c}$ being renormalized due to the
photon-photon interaction:
\begin{equation} \label{13}
\begin{array}{l}
\displaystyle{%
  H_S=\sum_{k,j}\hbar\tilde{\omega}_k\,a_{kj}^+a_{kj}+E_0, %
}
\end{array}
\end{equation}
where $\tilde{\omega}_k=\tilde{c} k$. The correlation Hamiltonian
describing the interaction between the renormalized or ``dressed''
photons is chosen from the condition that the full Hamiltonian 
should be unchanged:
\begin{equation} \label{14}
\begin{array}{l}
\displaystyle{%
  H_C=\sum_{k,j}\hbar\big(\omega_k-\tilde{\omega}_k\big)a_{kj}^+a_{kj}+\sum_{k}\hbar\omega_k - E_0 + H_I. %
}
\end{array}
\end{equation}
This Hamiltonian describes the interaction between photons
propagating with the renormalized speed of light, which we will not
consider. Formulas (\ref{13}),\,(\ref{14}) contain the non-operator
term $E_0$, taking account of which proves to be important for
correct formulation of the self-consistent field model. %
Let us choose it from the consideration that the approximating
Hamiltonian (\ref{13}) should be maximally close to the exact
Hamiltonian. This means we have to require that the quantity %
$I\equiv\big|\big\langle H-H_S\big\rangle\big|=\big|\big\langle H_C\big\rangle\big|$ %
should be minimal, that is equal to zero. From here we obtain the
conditions being natural for the self-consistent field theory:
\begin{equation} \label{15}
\begin{array}{l}
\displaystyle{%
  \big\langle H\big\rangle=\big\langle H_S\big\rangle, \quad \big\langle H_C\big\rangle = 0. %
}
\end{array}
\end{equation}
The averaging is performed by means of the statistical operator
\begin{equation} \label{16}
\begin{array}{l}
\displaystyle{%
  \rho=\exp\beta\big( F-H_S\big), %
}
\end{array}
\end{equation}
where $F$ is the free energy, $\beta=1/T$ is the inverse
temperature. The condition (\ref{15}) allows to determine the
non-operator part of the Hamiltonian (\ref{13}):
\begin{equation} \label{17}
\begin{array}{l}
\displaystyle{%
 E_0=2(c-\tilde{c})\sum_{k}\hbar k f_k + \sum_k \hbar ck\, +\big\langle H_I\big\rangle, %
}
\end{array}
\end{equation}
where the distribution function of the renormalized photons has the
Planck form
\begin{equation} \label{18}
\begin{array}{l}
\displaystyle{%
 f_k=\big\langle a_{kj}^+a_{kj}\big\rangle = \frac{1}{\exp(\beta\hbar\tilde{\omega}_k)-1} %
}
\end{array}
\end{equation}
and does not depend on the polarization index. From the
normalization condition for the statistical operator (\ref{16})
$\rm{Sp}\,\rho =1$ it follows the expression for the free energy of
radiation
\begin{equation} \label{19}
\begin{array}{l}
\displaystyle{%
 F=2(c-\tilde{c})\sum_{k}\hbar k f_k + \sum_k \hbar ck\, +\big\langle H_I\big\rangle\, + %
}\vspace{2mm}\\ %
\displaystyle{\hspace{06mm}%
  +\,2T\sum_k\ln\!\left( 1-e^{-\beta\hbar\tilde{\omega}_k}\!\right). %
}
\end{array}
\end{equation}
With neglect of the photon-photon interaction and zero fluctuations,
from formula (\ref{19}), of course, there follow the usual formulas
of the thermodynamics of blackbody radiation \cite{LL}. %
It is natural to require that in the used approximation with the
Hamiltonian (\ref{13}) and the free energy (\ref{19}), like in the
case of a gas of noninteracting photons, the thermodynamic relations
should hold. Since the introduced renormalized speed $\tilde{c}$
itself can, in principle, depend on thermodynamic variables, then in
order for the thermodynamic relations to hold the following
condition should be satisfied:
\begin{equation} \label{20}
\begin{array}{l}
\displaystyle{%
 \frac{\partial F}{\partial \tilde{c}} = 0.  %
}
\end{array}
\end{equation}
From this condition and formula (\ref{19}) it follows the relation
which determines the renormalized speed:
\begin{equation} \label{21}
\begin{array}{l}
\displaystyle{%
 \tilde{c}-c= \frac{ \displaystyle{\frac{\partial \big\langle H_I\big\rangle}{\partial \tilde{c}}} }
              {\displaystyle{2\frac{\partial}{\partial \tilde{c}}\sum_k \hbar kf_k} }.  %
}
\end{array}
\end{equation}
Since formula (\ref{21}) contains the temperature-dependent
distribution function (\ref{18}) then, naturally, also the speed of
light $\tilde{c}=\tilde{c}(T)$ is a function of temperature. Thus,
we have to calculate the average of the interaction Hamiltonian
$\big\langle H_I\big\rangle$. Here, as in the theory of phonons in
solids \cite{P6,P7}, divergent integrals appear. While describing
phonons within the continuum model it is natural to cut off such
integrals at the wave number, which equals the inverse average
distance between particles or, at integration over frequencies, at
the Debye frequency. In the case of photons, we will cut off
divergent integrals at some wave number $k_m$, the choice of which
is discussed a little later. With this in mind, the calculation of
the average of the interaction Hamiltonian (\ref{07}) gives
\begin{equation} \label{22}
\begin{array}{l}
\displaystyle{%
 \big\langle H_I\big\rangle=\frac{1312 V}{15\pi^2}D\,\hbar^2c^2J\left(\frac{k_m^4}{4}+J\right),  %
}
\end{array}
\end{equation}
where $\displaystyle{J=6\zeta(4)\left(\frac{T}{\hbar\tilde{c}}\right)^{\!4}}$, %
$\zeta(4)=\pi^4/90\approx 1.0823$ is the zeta function. Let
$\sigma\equiv \tilde{c}/c$ be the ratio of the temperature-dependent
speed of light to the ``bare'' speed of light. Considering that
$\displaystyle{\sum_k \hbar kf_k=\frac{V\hbar}{2\pi^2}J}$, from
(\ref{21}) we get the equation for $\sigma$:
\begin{equation} \label{23}
\begin{array}{l}
\displaystyle{%
 \sigma=1+\frac{328}{15}D\hbar c\, k_m^4 + \frac{328\cdot\!16}{5}\zeta(4)D\hbar c\left(\frac{T}{\hbar c}\right)^{\!4}\!\cdot\!\frac{1}{\sigma^4}. %
}
\end{array}
\end{equation}
This implies that the ratio of the speed of light at zero
temperature $\tilde{c}_0$ to the ``bare'' speed of light
$\sigma_0\equiv \tilde{c}_0/c$ is determined by the formula:
\begin{equation} \label{24}
\begin{array}{l}
\displaystyle{%
 \sigma_0 =1+\frac{328}{15}D\hbar c\, k_m^4. %
}
\end{array}
\end{equation}
It is the speed of light at zero temperature that is a directly
measurable speed. As follows from (\ref{24}), this speed does not
coincide with the ``bare'' speed of light, which is caused by taking
into account the interaction between photons. Because of the
weakness of this interaction $c$ and $\tilde{c}_0$ should differ
very little and in the main approximation they could be considered
equal, which would not affect further conclusions. Nevertheless, it
is of certain interest to clarify in more detail the relation
between $c$ and $\tilde{c}_0$, which, as seen from (\ref{24}), is
essentially determined by the choice of the wave number $k_m$ at
which the cut-off of divergent integrals is carried out. %
We find this wave number from the condition $\hbar\tilde{c}_0k_m=m\tilde{c}_0^2$, %
so that $k_m$ is equal to the inverse Compton wavelength of an
electron $k_m=m\tilde{c}_0/\hbar=\lambdabar_0^{-1}$.  %
This condition implies that a real electron cannot be created from
the energy of zero oscillations. A similar method of cutting off
divergent integrals was employed, for example, by Bethe in the
nonrelativistic calculation of the Lamb shift \cite{Bethe}. %
With such cut-off procedure, from (\ref{24}) it follows
\begin{equation} \label{25}
\begin{array}{l}
\displaystyle{%
 \sigma_0 =1+\chi\sigma_0^6, %
}
\end{array}
\end{equation}
where $\displaystyle{\chi\equiv\frac{328}{15}\eta_0}$,
$\displaystyle{\eta_0\equiv\frac{\alpha_0^2}{45(4\pi)^2}}$ and
$\displaystyle{\alpha_0\equiv\frac{e^2}{\hbar \tilde{c}_0}}$ is the
fine-structure constant written through the observed speed of light.
Formula (\ref{25}) determines the ratio $\sigma_0\equiv \tilde{c}_0/c$ %
through the observed fine-structure constant. With the help of it,
the unobserved ``bare'' speed can be eliminated from Eq.\,(\ref{23}). %
As a result, we come to the equation for the dimensionless quantity
$\tilde{\sigma}\equiv\sigma/\sigma_0=\tilde{c}/\tilde{c}_0$, %
which equals the ratio of the observed speeds of light at finite and
at zero temperatures:
\begin{equation} \label{26}
\begin{array}{l}
\displaystyle{%
 \tilde{\sigma}^5-\tilde{\sigma}^4 = b\tau^4. %
}
\end{array}
\end{equation}
Here $b\equiv 3\chi\sigma_0^5\approx 4.9\cdot\!10^{-7}$, $\tau\equiv
T/T_0$ is the dimensionless temperature, and $T_0$ is characteristic
temperature determined by the rest energy of an electron
\begin{equation} \label{27}
\begin{array}{l}
\displaystyle{%
  m\tilde{c}_0^2=2 [\zeta(4)]^{1/4}T_0, %
}
\end{array}
\end{equation}
so that $T_0\approx 0.29\cdot\!10^{10}$\,K. Thus, it follows from
formula (\ref{26}) that the speed of light rises with increasing
temperature. As opposed to the ``bare'' photons with the dispersion
law $\omega=ck$, the photons which speed is determined by the
self-consistency Eq.\,(\ref{26}) and depends on temperature have the
dispersion law $\tilde{\omega}=\tilde{c}k$, and it is natural to
call them ``self-consistent'' photons.

At $b\tau^4\ll 1$ we have $\tilde{\sigma}\approx 1+b\tau^4$. Since
the coefficient $b$ is very small, then the temperature dependence
of the speed of light can manifest itself only at very high
temperatures. For the observed relict radiation with the temperature
$T=2.73$\,K we have $\tilde{\sigma}-1\approx 3.8\cdot\!10^{-43}$, so
that the speed of light practically coincides with the speed of
light at zero temperature. Inside stars, temperature can reach tens
of millions degrees. For example, at the temperature inside the Sun
that equals 15 million degrees, we have $\tilde{\sigma}-1\approx
3.4\cdot\!10^{-16}$. This means that the speed of light inside the
Sun differs from the speed of light at zero temperature by the
amount $\Delta\tilde{c}=\tilde{c}-\tilde{c}_0\approx 10^{-5}\,$cm/s.
In order for the speed of light of the equilibrium radiation at a
finite temperature to differ from the speed of light at zero
temperature by one percent $\tilde{\sigma}=1.01$, the temperature
$T\approx 12\,T_0\approx 3.5\cdot\!10^{10}\,$K is required. Note
that the temperature correction to the speed of light obtained above
is proportional to $T^4$, as in work \cite{Tarrach}, but has the
opposite (positive) sign. This is possibly caused by that the
influence of the mean field, created by all photons, is taken into
account in the present approach.

Although, as was noted, the Hamiltonian (\ref{03}) is valid at low
frequencies and, therefore, low temperatures, nevertheless,
considering this Hamiltonian as a model one, it is of interest to
consider also the case of high temperatures. As it turns out, in
this limit we also obtain consistent results, in particular all
thermodynamic relations prove to be satisfied. Therefore, there is
reason to hope that the obtained results will, at least
qualitatively, correctly describe the influence of the interaction
of photons on the speed of light in this high-temperature limit.

In the limit of very high temperatures $\tau\gg b^{-1/4}\approx 38$
we have
\begin{equation} \label{28}
\begin{array}{l}
\displaystyle{%
  \tilde{\sigma}\approx b^{1/5}\tau^{4/5}. %
}
\end{array}
\end{equation}
In the modern epoch this condition is apparently not realized. But
accounting for the dependence of the speed of light on temperature
should be of principal importance in the very early stage of
evolution of the Universe, when the dependence (\ref{28}) could be
valid. In the model of the hot Universe \cite{ZN}, in the first
instances after the Big Bang the temperature of the Universe was
anomalously high in comparison with modern temperatures. As follows
from the relations obtained above, also the speed of light was large
in comparison with the present one. As the Universe was expanding
and cooling the speed of light was decreasing and in the modern
epoch it reached its value, practically equal to that of the speed
of light at zero temperature. At the Planck temperature
$T_p\approx 1.42\cdot\!10^{32}\,\rm{K}\approx 10^{19}\,\rm{GeV}$ %
the speed of light $\tilde{c}_p$ had to exceed the present one by
many orders of magnitude: $\tilde{c}_p/\tilde{c}_0\approx 0.8\cdot\!10^{17}$. %
The illustration of how the speed of light was varying as the
Universe was cooling in the first instances after the Big Bang is
given in Table I.
\vspace{-04mm}%
\begin{table}[h!] \nonumber
\centering %
\caption{The value of the speed of light at different temperatures in the first instances after the Big Bang} %
\vspace{0.5mm}%
\begin{tabular}{|c|c|c|c|c|c|} \hline  
$t,s$                   & $T$,\,GeV               & $T$,\,K                  & $\tau=T/T_0$             & $n$,\,cm$^{-3}$        &  $\tilde{c}/\tilde{c}_0$   \\ \hline %
$5.4\!\cdot\!10^{-44}$  & $1.2\!\cdot\!10^{19}$   & $1.42\!\cdot\!10^{32}$   & $4.9\!\cdot\!10^{22}$    & $1.3\!\cdot\!10^{47}$  &  $0.8\!\cdot\!10^{17}$     \\ \hline %
$10^{-39}$              & $10^{16}$               & $10^{29}$                & $3.5\!\cdot\!10^{19}$    & $1.6\!\cdot\!10^{45}$  &  $2.3\!\cdot\!10^{14}$     \\ \hline %
$10^{-11}$              & 100                     & $10^{15}$                & $3.5\!\cdot\!10^{5}$     & $6.5\!\cdot\!10^{36}$  &  $1.5\!\cdot\!10^{3}$      \\ \hline %
$10^{-5}$               & 0.2                     & $2\!\cdot\!10^{12}$      & $6.9\!\cdot\!10^{2}$     & $1.4\!\cdot\!10^{35}$  &  $10$                      \\ \hline %
$10^{-2}$               & $10^{-2}$               & $2\!\cdot\!10^{11}$      & $69$                     & $2.5\!\cdot\!10^{34}$  &  $1.9$                     \\ \hline %
$1.5$                   & \,$0.7\!\cdot\!10^{-3}$ & $0.8\!\cdot\!10^{10}$    & $2.8$                    & $4.9\!\cdot\!10^{30}$  &  $1.00003$                 \\ \hline %
\end{tabular}  
\end{table}

\noindent The reason of the large effect at small times at weak
photon-photon interaction, as seen from the second-to-last column of
the table, is the extremely large density of photons at such
temperatures.

\section{Thermodynamics of the equilibrium radiation of self-consistent photons}\vspace{-0mm}
Let us give general formulas for the thermodynamic functions of a
gas of self-consistent photons. The free energy (\ref{19}),
expressed through the observed speed of light, can be written in the form %
\begin{equation} \label{29}
\begin{array}{l}
\displaystyle{%
  \frac{F}{U_V}= 1+3\sigma_0\left(1-\frac{4}{3}\,\tilde{\sigma}\right)\frac{\tau^4}{\tilde{\sigma}^4} %
  +\frac{9}{2}\,\chi\sigma_0^6\,\frac{\tau^8}{\tilde{\sigma}^8},
}
\end{array}
\end{equation}
where
$\displaystyle{U_V\equiv\frac{V}{8\pi^2}\frac{m\tilde{c}_0^{\,2}}{\lambdabar_0^3}\,\sigma_0^{-1}}$ %
is the energy of zero oscillations. It is easy to verify that Eq.
(\ref{26}) follows from the condition
$\displaystyle{\frac{\partial}{\partial\tilde{\sigma}}\left(\frac{F}{U_V}\right)=0}$. %
This condition allows to calculate from the expression for the free
energy (\ref{29}) by the usual formulas the pressure
$\displaystyle{p=-\left(\frac{\partial F}{\partial V}\right)_{\!T}}$ and the entropy %
$\displaystyle{S=-\left(\frac{\partial F}{\partial
T}\right)_{\!V}}$, and due to fulfillment of this condition the
temperature-dependent parameter $\tilde{\sigma}$ should not be differentiated. %
Considering (\ref{26}), the formulas for the pressure and entropy can be written in the form %
\begin{equation} \label{30}
\begin{array}{l}
\displaystyle{%
  p=-\frac{U_V}{V}\!\left[1+\frac{3\sigma_0}{2}\left(\!1-\frac{5}{3}\,\tilde{\sigma}\!\right)\!\frac{\tau^4}{\tilde{\sigma}^4}\right], %
}
\end{array}
\end{equation}
\begin{equation} \label{31}
\begin{array}{l}
\displaystyle{%
  S=\frac{4\sigma_0 U_V}{T_0}\frac{\tau^3}{\tilde{\sigma}^3}. %
}
\end{array}
\end{equation}
The total energy $E=F+TS$ is
\begin{equation} \label{32}
\begin{array}{l}
\displaystyle{%
  E=U_V\!\left[1+\frac{3}{2}\,\sigma_0\big(1+\tilde{\sigma}\big)\frac{\tau^4}{\tilde{\sigma}^4}\right]. %
}
\end{array}
\end{equation}
With neglect of the interaction between photons, when $\sigma_0=\tilde{\sigma}=1$, %
and without taking into account vacuum fluctuations formulas
(\ref{29})\,--\,(\ref{32}) turn into classical formulas of the
theory of blackbody radiation \cite{LL}. In order to pass to this
limit it is convenient to use the formula
$\displaystyle{\frac{U_V}{T_0^4}=V\frac{2\zeta(4)}{\pi^2\sigma_0\hbar^3\tilde{c}_0^{\,3}}=V\frac{\pi^2}{45\sigma_0\hbar^3\tilde{c}_0^{\,3}}}$. %
But, even with neglect of the interaction between photons,
accounting for vacuum fluctuations leads to appearance of the
additional energy $U_V$ in the total energy of blackbody radiation,
and a negative contribution from vacuum fluctuations appears in the
expression for the pressure (\ref{30}). Hence, instead of the usual
relation between energy and pressure $pV=E/3$ \cite{LL}, with
account of fluctuations we obtain $3pV=E-4U_V$. At temperatures
$T<T_0$ the total pressure proves to be negative and changes sign,
becoming positive, at $T>T_0$. Accounting for the interaction
between photons leads to a little shift of the temperature at which
the pressure changes sign. This temperature $T_1$ can be found from
Eqs. (\ref{26}) and (\ref{30}), that gives $T_1\approx T_0(1+3\chi)$. %
Note that vacuum fluctuations do not give a contribution into the
enthalpy
\begin{equation} \label{33}
\begin{array}{l}
\displaystyle{%
  W=E+pV=4U_V\sigma_0\frac{\tau^4}{\tilde{\sigma}^3}, %
}
\end{array}
\end{equation}
as well as into the entropy (\ref{31}).
For calculation of the heat capacity of a gas of photons %
$C_V=T\big(\partial S/\partial T\big)_V$, %
it is already necessary to account for the dependence of the speed
of light on temperature, using the formula (\ref{26}), so that we
obtain
\begin{equation} \label{34}
\begin{array}{l}
\displaystyle{%
  C_V=\frac{12\sigma_0 U_V}{T_0}\frac{\tilde{\sigma}}{(5\tilde{\sigma}-4)}\frac{\tau^3}{\tilde{\sigma}^3}. %
}
\end{array}
\end{equation}
Let us also give the formula for the number of photons
\begin{equation} \label{35}
\begin{array}{l}
\displaystyle{%
  N=\frac{90\zeta(3)}{\pi^4}\frac{U_V}{T_0}\sigma_0\frac{\tau^3}{\tilde{\sigma}^3}. %
}
\end{array}
\end{equation}
In the low-temperature limit $T\ll T_0$ formulas
(\ref{33})\,--\,(\ref{35}), of course, turn into the known formulas
of the theory of blackbody radiation \cite{LL}.

\section{The photon distribution function and thermodynamics of
the equilibrium radiation at high temperatures}\vspace{0mm} %
Despite the fact that description of the interaction of photons is
based on the Euler-Heisenberg Hamiltonian (\ref{03}), which is
strictly speaking valid for sufficiently low frequencies, however
the general formulas obtained by means of it lead to correct and
consistent thermodynamic relations also in the limit of high
temperatures. In the region of high temperatures $\tau\gg b^{-1/4}$,
the distribution functions for the number of photons and the energy
with respect to wave numbers have the Planck form at all
temperatures
\begin{equation} \label{36}
\begin{array}{l}
\displaystyle{%
  n_k=\frac{k^2}{\pi^2\big(e^{Lk}-1\big)}, \qquad %
  \varepsilon_k=\frac{\hbar\tilde{c}k^3}{\pi^2\big(e^{Lk}-1\big)}, %
}
\end{array}
\end{equation}
where $L\equiv\hbar\tilde{c}/T$, so that the total densities of the
number of photons and energy are respectively
$n=\int_0^\infty\!n_kdk$ and $\varepsilon=\int_0^\infty\!\varepsilon_kdk$. %
However, the parameter $L$ entering into (\ref{36}) depends on
temperature differently in the low-temperature and high-temperature limits. %
At low temperatures $L\equiv\hbar\tilde{c}_0/T=\big[2\pi\big/90^{1/4}\big]\!(\lambdabar_0/\tau)$, %
and at high temperatures $L=B\lambdabar_0\big/\tau^{1/5}$, where
$B=2\pi b^{1/5}\big/90^{1/4}=0.11$. In particular, in the case of
low temperatures, as is known, maximums of the distributions
(\ref{36}) shift to higher energies proportional to temperature,
respectively as $k_{\rm{max}}\lambdabar_0=0.782\,\tau$ and
$k_{\rm{max}}\lambdabar_0=1.383\,\tau$ (Wien's displacement law) \cite{LL}. %
In the limit of high temperatures $\tau\gg b^{-1/4}$ maximums of the
distributions (\ref{36}) also shift to higher energies with
increasing temperature, but much slower, as
$k_{\rm{max}}\lambdabar_0=14.51\,\tau^{1/5}$ and
$k_{\rm{max}}\lambdabar_0=25.65\,\tau^{1/5}$ respectively.

When going over to the distribution functions with respect to frequencies %
$\tilde{\omega}=\tilde{c}k$ in (\ref{36}) $Lk\equiv\hbar\tilde{\omega}\big/T$, %
and maximums of the distributions with respect to frequencies %
shift with temperature in the same way at all temperatures: as
$\hbar\tilde{\omega}_{\rm{max}}\big/T=1.594$ for the number of photons and %
$\hbar\tilde{\omega}_{\rm{max}}\big/T=2.821$ for the energy. 

In the high-temperature limit the temperature dependencies of
thermodynamic functions of the equilibrium radiation are determined by the formulas: %
\begin{equation} \label{37}
\begin{array}{ccc}
\displaystyle{\hspace{0mm}%
  p=\frac{5}{2}\frac{U_V\sigma_0}{V}\frac{\tau^{8/5}}{b^{3/5}},\qquad E=\frac{3}{2}\,U_V\sigma_0\frac{\tau^{8/5}}{b^{3/5}}, %
}\vspace{2mm}\\ %
\displaystyle{\hspace{0mm}%
  S=\frac{4U_V\sigma_0}{T_0}\left(\frac{\tau}{b}\right)^{\!3/5},\qquad C_V=\frac{12U_V\sigma_0}{5T_0}\left(\frac{\tau}{b}\right)^{\!3/5}, 
}\vspace{2mm}\\ %
\displaystyle{\hspace{0mm}%
  W=4U_V\sigma_0\frac{\tau^{8/5}}{b^{3/5}},\qquad N=\frac{90\zeta(3)}{\pi^4}\frac{U_V\sigma_0}{T_0}\left(\frac{\tau}{b}\right)^{\!3/5}. %
}
\end{array}
\end{equation}
Here the pressure and the energy are connected by the relation
$\displaystyle{pV=\frac{5}{3}\,E}$. As seen, the thermodynamic
quantities increase with temperature much slower than at low
temperatures. Note that the calculation of the number density of
photons in Table I is made by means of formula (\ref{37}), whereas
the calculation by the standard formula of the theory of blackbody
radiation gives even more high density of photons.

\section{Conclusion}\vspace{-0mm}
The considered effect of dependence of the speed of light in vacuum
on the radiation temperature is of fundamental importance for
understanding the world around us and the early stage of evolution
of the Universe. In the theories of special and general relativity
the speed of light in vacuum is considered to be a cosmological
constant. The equations of Einstein's theory of general relativity
are usually written in such form that their left part is expressed
through the space-time curvature tensor and has a purely geometric
nature, and the right part contains the energy-momentum tensor of
matter and fields of different nature. As is known, Einstein himself
was dissatisfied with such separation of geometry and matter in the
equations. Accounting for dependence of the speed of light on
conditions, at which its propagation occurs, results in that now the
metric tensor itself, through the speed of light contained in it,
proves to be directly dependent on the state of matter, and thus the
interdependence of matter and geometry becomes closer.

Einstein was rather interested in evidence for the possible dependence
of the speed of light on the external conditions. %
As P.L.\,Kapitsa recalled \cite{Kapitsa}, when,
working in the 30s of past century in the Cavendish Laboratory with Rutherford, %
he obtained magnetic fields 10 times stronger than those obtained before, %
a number of scientists advised him to make experiments on studying
the influence of strong magnetic field on the speed of light. The
one who insisted the most was Einstein. He said to Kapitsa: ``I
don't believe that God created such the Universe, that the speed of
light depends on nothing in it''. Yet Kapitsa refused the proposed
experiment, on the ground that the experiment promised to be
extremely difficult and the effect, if it had been discovered, for
sure would have been at the edge of experimental accuracy and there
would have been no credit to these results.

The above-stated calculations of dependence of the speed of light on
temperature allow to definitely conclude that, as Einstein surmised, %
the magnetic field, similarly to temperature, will affect the speed
of light propagation. We can estimate the order of magnitude of
fields, at which the speed of light will change substantially, by
equating the energies (\ref{02}) and (\ref{03}) $w_0\sim w_I$. The
estimation gives $H\sim 10^{16}$\,Gs. So, P.L.\,Kapitsa was right
when he refused to perform a labor-consuming experiment, because the
fields necessary for observation of such effect should be so strong
that they could hardly be realized in modern conditions.

The performed calculations are based on the Hamiltonian (\ref{03}),
the use of which is valid at sufficiently low frequencies and
temperatures. However, the obtained formulas lead to reasonable
relations also at high temperatures, so we can hope that they, at
least qualitatively, correctly reflect the real situation. To make
more consistent and strict calculation one should reformulate the
quantum electrodynamics, choosing not the free electron-positron and
photon fields as the main approximation, but the Hamiltonian in the
self-consistent field model, in a similar way as, for example, it
was done for the scalar and Dirac fields in \cite{P4,P5}. This is,
certainly, a rather complex problem.

So long as, according to above mentioned estimates, in the first
instances of existence of the Universe the speed of light exceeded
its present value by many orders of magnitude, this should
substantially affect the existing scenarios of the evolution of the
Universe at its early stage.


\end{document}